# Carbon Nanotubes as Electrodes for Dielectrophoresis of DNA

*Sampo Tuukkanen, J. Jussi Toppari, Anton Kuzyk, Lasse Hirviniemi, Vesa P. Hytönen[‡], Teemu Ihalainen[†], and Päivi Törmä\**

Nanoscience Center, Department of Physics and [†]Department of Biological and Environmental Science, University of Jyväskylä, P.O.Box 35 (YN), FIN-40014 Jyväskylä, Finland, and [‡]Department of Materials, ETH Zürich, Hönggerberg, CH-8093 Zürich, Switzerland.

\*email: paivi.torma@phys.jyu.fi. FAX: +358 14 260 4756. TEL: +358 14 260 2384.

Dielectrophoresis can potentially be used as an efficient trapping tool in the fabrication of molecular devices. For nanoscale objects, however, the Brownian motion poses a challenge. We show that the use of carbon nanotube electrodes makes it possible to apply relatively low trapping voltages and still achieve high enough field gradients for trapping nanoscale objects, e.g., single molecules. We compare the efficiency and other characteristics of dielectrophoresis between carbon nanotube electrodes and lithographically fabricated metallic electrodes, in the case of trapping nanoscale DNA molecules. The results are analyzed using finite element method simulations and reveal information about the frequency dependent polarizability of DNA.

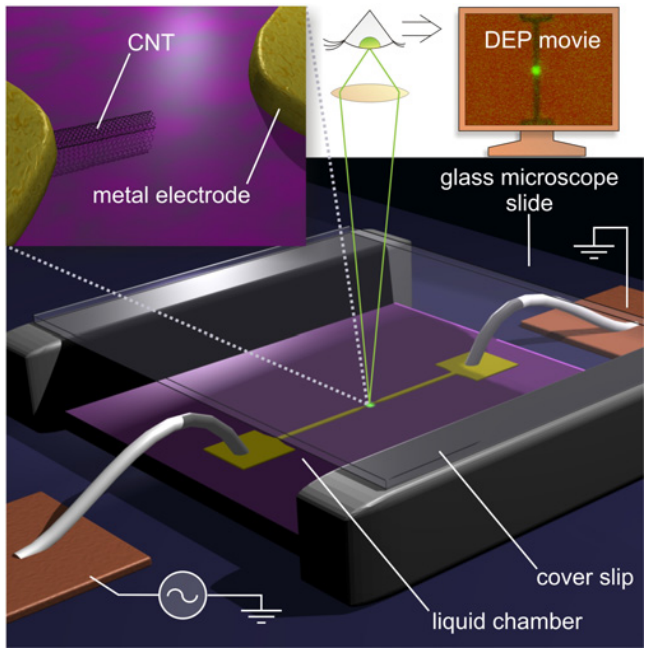

**Figure 1.** A schematic view of the experimental setup used in the DEP experiments under the confocal microscope. The solution containing DNA is in the moisture chamber between the silicon substrate and the cover slip. The structure with carbon nanotube as one electrode is presented in the close-up image. Repeated confocal microscope images are captured to obtain time-resolved information about the DEP process (the "DEP movie").

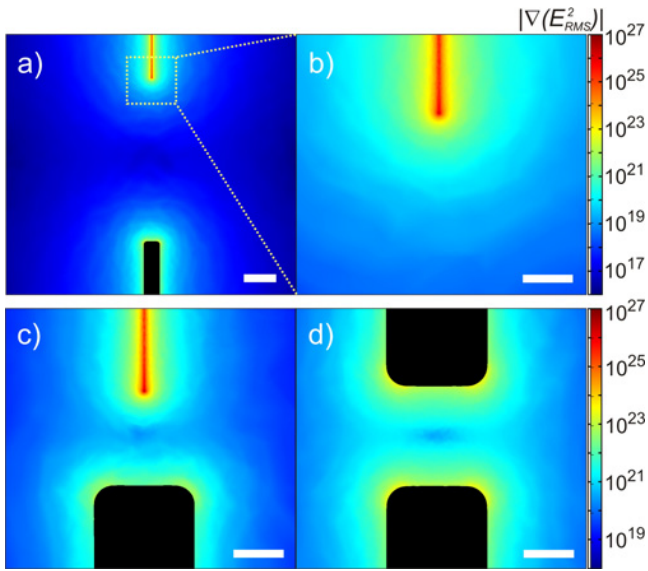

**Figure 2.** Contour plot of the gradient of the field square, $\nabla(E_{rms}^2)$, in the plane 2 nm above the substrate surface, i.e., 0.2 nm above the CNT in (a–c). The DEP force has the maximum value in the very end of the CNT in (a–c). In (a), and the close-up (b), the gap size is 1 μm and in (c) and (d) the gap size is 100 nm. The dc voltage between the electrodes is 1.6 $V_{rms}$. The scale bars are 200 nm in (a) and 50 nm in (b–d).

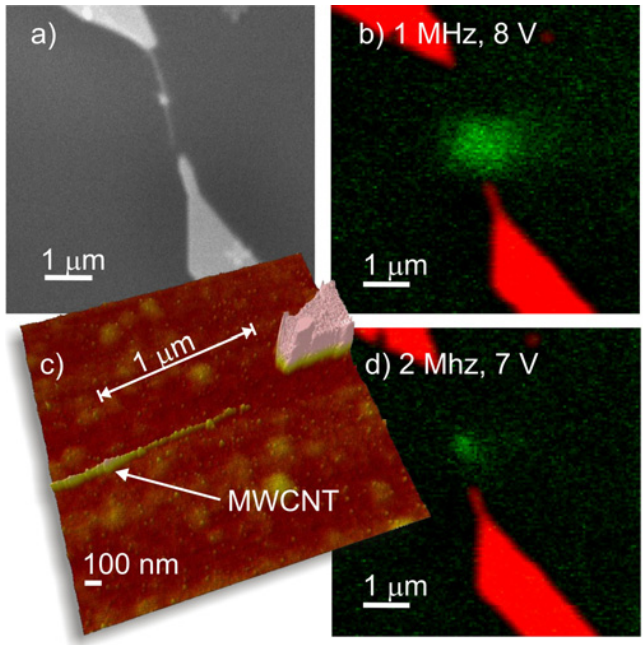

**Figure 3.** DEP of 1065 bp dsDNA using CNT as one electrode: (a) SEM and (c) AFM images of the multiwalled CNT electrode sample before confocal experiment. (b) and (d) show the trapped DNA spot when a certain frequency and voltage were used.

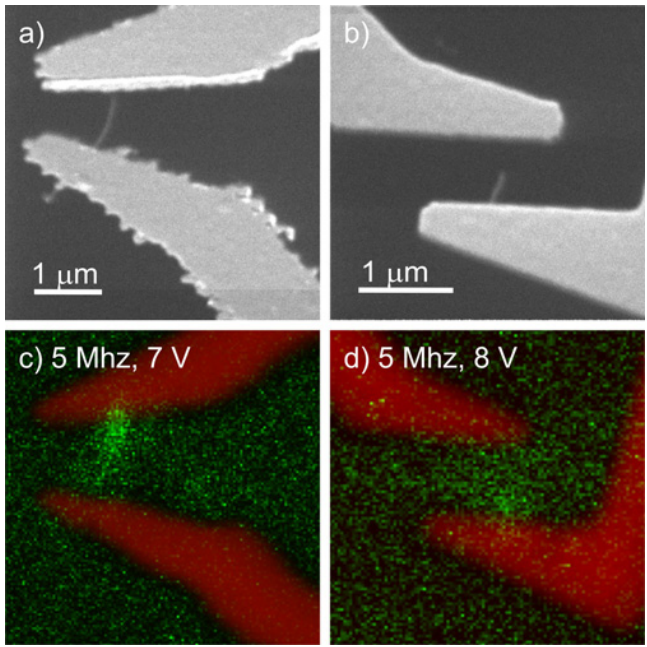

**Figure 4.** DEP of 145 bp dsDNA using CNT as an electrode. (a) and (b) are SEM images of the CNT electrode samples and (c) and (d) are corresponding fluorescence images taken during the DEP, using the shown frequency and voltage. The gap sizes are in (a) ~115 nm and in (b) ~350 nm.

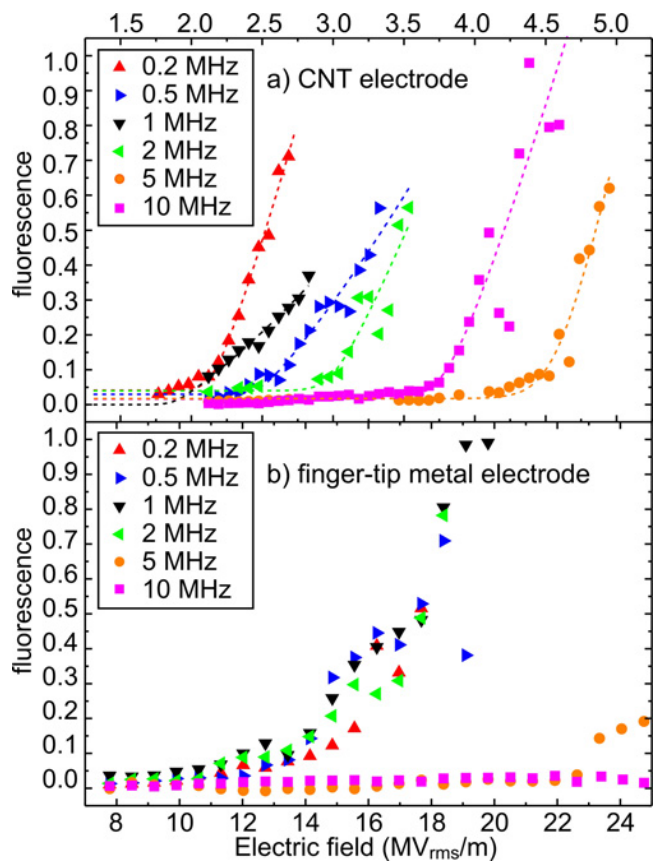

**Figure 5.** Comparison of the trapping efficiency of CNT electrode vs finger-tip electrodes. The curves show the fluorescence (a) in the end of CNT (with electrode separation $d = 1$ μm) and (b) in the gap (in the case of finger-tip metal electrodes separation $d = 100$ nm) as a function of the electric field (an average electric field strength between the electrodes, $E = V/d$). Dotted lines in (a) are fits to the data using the function $I = I_0 + A(V^b + V_{min}^b)^{2/b}$ (see text). By comparing the field strength needed to trap DNA in these cases, one can clearly see that CNT electrode shows better performance than lithographically fabricated nanoelectrodes.

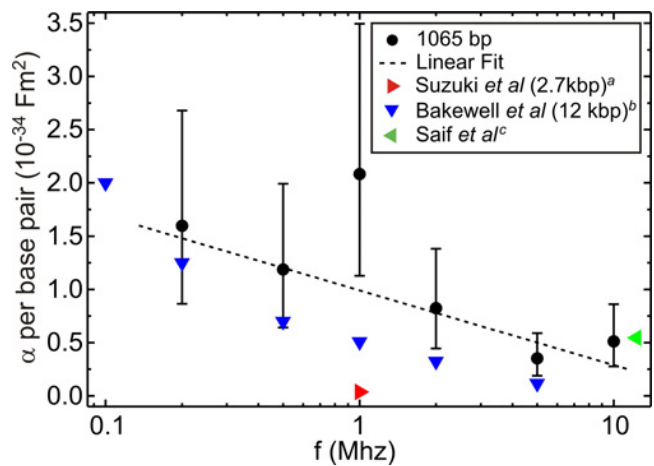

**Figure 6.** Polarizability of 1065 bp DNA calculated from the fluorescence data captured during DEP using CNT electrode sample shown in Figure 3. The error bars originate from the uncertainty of the observed fluorescence spot radius (0.5 ± 0.1 μm). Other values are taken from refs [a]20, [b]23, and [c]24.